\documentclass[a4paper,pre,twocolumn,superscriptaddress,showkeys,floatfix]{revtex4-1}
\usepackage[dvips]{graphicx}
\usepackage{amsmath}
\usepackage{amsthm}
\usepackage{amssymb}

\newcommand{\mean}[1]{\left\langle #1 \right\rangle}
\newcommand{\var}[1]{\mathrm{var}\!\left( #1 \right)}

\newcommand{\erg}{\mathcal U}

\begin{document}

\title{The scaling window of the 5D Ising model with free boundary conditions}

\date{\today} 

\author{P. H. Lundow} 
\email{per.hakan.lundow@math.umu.se} 

\author{K. Markstr\"om}
\email{klas.markstrom@math.umu.se} 

\affiliation{ Department of mathematics and mathematical statistics,
  Ume\aa{} University, SE-901 87 Ume\aa, Sweden}

\begin{abstract}
  The five-dimensional Ising model with free boundary conditions has
  recently received a renewed interest in a debate concerning the
  finite-size scaling of the susceptibility near the critical
  temperature.  We provide evidence in favour of the conventional
  scaling picture, where the susceptibility scales as $O(L^2)$ inside
  a critical scaling window of width $O(1/L^2)$. Our results are based
  on Monte Carlo data gathered on system sizes up to $L=79$ (ca. three
  billion spins) for a wide range of temperatures near the critical
  point. We analyse the magnetisation distribution, the susceptibility
  and also the scaling and distribution of the size of the
  Fortuin-Kasteleyn cluster containing the origin.  The probability of
  this cluster reaching the boundary determines the correlation
  length, and its behaviour agrees with the mean field critical
  exponent $\delta=3$, that the scaling window has width $O(1/L^2)$.
\end{abstract}


\maketitle

\section{Introduction}
The Ising model in dimension $d=5$ is of particular interest since it
is the first case where the model is strictly above its upper critical
dimension $d_c=4$.  Rigorous results \cite{aizenman:82, sokal:79}
establish that the critical exponents of the model assume their mean
field values. Here the specific heat exponent $\alpha=0$, and the
results of \cite{sokal:79} also imply that the specific heat is
bounded at the critical point.  Recent simulation results
\cite{lundow:15a} also indicate that, just as for the mean field case,
the specific heat is discontinuous at the critical point.

In contrast to these asymptotic results there has been a long running
debate over the finite size scaling for the model with free boundary
conditions. We'll refer the reader to \cite{berche:12} for a fuller
overview of the history and stick to the presently most relevant parts
here. For $d=5$ and cyclic boundary conditions there is agreement that
e.g. $\chi \propto L^{5/2}$ for a lattice of side $L$.  The
conventional picture for the free boundary case is that $\chi \propto
L^{2}$. However, it has also been suggested \cite{berche:12}
that the free boundary case should scale in the same way as the cyclic
boundary case near the finite size susceptibility maximum.  A computational 
study \cite{boundarypaper} of the,
then, largest lattices possible supported the conventional picture,
but in \cite{berche:12} it was suggested this was due to an
underestimate of the influence of the large boundaries of the used
systems.  For systems exactly at the critical coupling this issue was
resolved in \cite{lundow:14} where a study of systems up to
$L=160$ demonstrated an increasingly good agreement with the
conventional picture as the system size was increased. But, this left 
the behaviour in the rest of the critical scaling window open.

The aim of the current paper is to extend the study of large systems
from \cite{lundow:14} to the full critical window, including the
location of the maximum of $\chi$, and give the best possible estimates
for the scaling behaviour in the coupling region discussed by all the
previous papers.  Apart from the susceptibility we also study
properties of the Fortuin-Kasteleyn cluster containing the origin  and 
use those to estimate both the susceptibility and the correlation length 
of the model.

To concretize, the predictions from \cite{berche:12} are that the
location of the maximum for $\chi$ will scale as $L^{-2}$ and the
maximum value as $L^{5/2}$.  The more recent \cite{WY2014} agrees 
with these predictions, and also the prediction from \cite{Ru85} that the 
location of the maximum of the susceptibility should scale as $L^2$, as does
\cite{Berche2015}, but both are based on smaller system sizes than 
those considered in the present work.

In short, our conclusion is that the data is well fitted by the
conventional scaling picture, both for the location and value of the
susceptibility, and location of the finite size critical point for the
magnetization.

\section{Definitions and details}
For a given graph $G$ on $N$ vertices the Hamiltonian with
interactions of unit strength along the edges is
$\mathcal{H}=-\sum_{ij} s_i s_j$ where the sum is taken over the edges
$ij$. Here the graph $G$ is a $5$-dimensional grid graph of linear
order $L$ with free boundary conditions, i.e. a cartesian product of
$5$ paths on $L$ vertices, so that the number of vertices is $N=L^5$
and the number of edges is $5L^5(1-1/L)$.  We use $K=1/k_BT$ as the
dimensionless inverse temperature (coupling) and denote the thermal
equilibrium mean by $\mean{\cdots}$. The critical coupling $K_c$ was
recently estimated by us~\cite{lundow:15a} to $K_c=0.11391498(2)$. We
will define a rescaled coupling as $\kappa=L^2(K-K_c)$ which gives a
scaling window of width $O(1/L^2)$.  The standard definitions apply;
the magnetisation is $M=\sum_i S_i$ (summing over the vertices $i$)
and the energy is $E=\sum_{ij}S_iS_j$ (summing over the edges
$ij$). We let $m=M/N$, $U=E/N$ and $\erg=\mean{U}$.

The susceptibility is $\chi=\mean{M^2}/N$ while we define the modulus
susceptibility as $\bar\chi=\var{|M|}/N$.  The standard deviation is
denoted $\sigma$, as is customary.  We will refer to the point where
the distribution of $M$ changes from unimodal to bimodal as
$K^*_c(L)$, or, in its rescaled form, $\kappa^*_c(L)$.  Recall also
that thermodynamic derivatives of e.g. $\log\chi$ can be obtained
through correlation measurements
$\partial\log\mean{|M|^{\ell}}/\partial K=\left(\langle E\,
|M|^{\ell}\rangle/\langle |M|^{\ell}\rangle\right) - \langle
E\rangle$.

Let $\langle S \rangle$ denote the average size of a flipped cluster
and recall~\cite{wolff:89} that $\chi=\langle S\rangle$. We use the
subscript $o$ to denote the origin (i.e. centre) vertex of the grid
$G$ so that $\langle S_o \rangle$ is the average size of a cluster
containing the origin.  The event that a cluster containing the origin
also reaches the boundary $\partial G$ of the graph is denoted
$o\leftrightarrow\partial G$.

We have collected data using Wolff-cluster spin updating for $L=15$,
$19$, $23$, $31$, $39$, $47$, $55$, $63$, $79$ for a wide and dense
set of $\kappa$-values in $0\le \kappa \le 0.92$. The data include
e.g. magnetisation, energy ($L\le 63$), size and radius of the origin
cluster. The total number of measurements range from about $300000$
for the small systems ($L\le 39$) down to $100000$ for $L=63$ and
$50000$ for $L=79$.

\section{The width of the scaling window and the shift exponents}
We begin by establishing the natural width of the scaling window,
i.e. the form of a natural rescaled temperature.  We have already
defined the rescaled coupling as $\kappa=L^2(K-K_c)$ with the
expectation that this will be the relevant scaling.  In principle, the
location of different effective finite size critical points could
scale with different exponents and we will compare a number of such
possibilities.

\begin{figure}
  \begin{center}
    \includegraphics[width=0.483\textwidth]{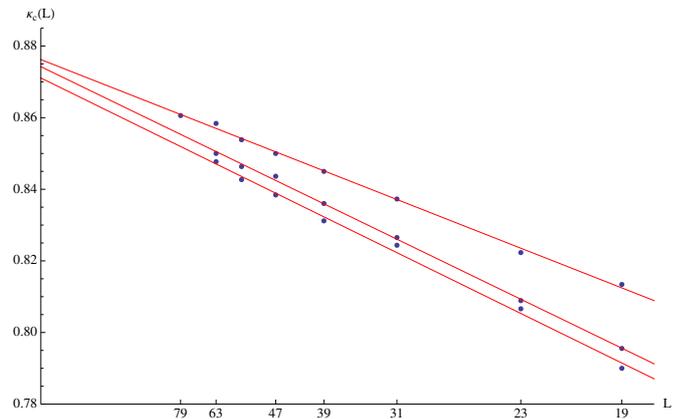}
  \end{center}
  \caption{(Colour on-line) Location $\kappa_c(L)$ of the maximum
    $\bar{\chi}$, $\partial\log\mean{|m|}/\partial K$ and
    $\partial\log\chi/\partial K$ plotted against $1/L$ for $L=19$,
    $23$, $31$, $39$, $47$, $55$, $63$ and $79$ ($L=79$ only for
    $\bar\chi$). Fitted lines are respectively $0.8763-1.21x$,
    $0.8742-1.49x$ and $0.8725-1.50x$ (downwards in figure) where
    $x=1/L$.}\label{fig:barchikcL}
\end{figure}

We used a 2nd order interpolation of the data points to estimate the
location of the maximum for three different parameters: $\bar{\chi}$,
$\partial\log\mean{|m|}/\partial K$ and $\partial\log\chi/\partial K$.
For the last two we only have these data for $L\le 63$.  The locations
are plotted versus $1/L$ in Figure~\ref{fig:barchikcL}.  The values
clearly point to three different limit $\kappa$-values.  Based on
$L\ge 19$ we estimate the limit values and an error bar by removing
each of the data points in turn and fitting a line to the remaining
points.  Thus we find $\kappa_c(L) = 0.8763(5) - 1.21(2)/L$ for the
$\bar\chi$-maximum, $\kappa_c(L) = 0.8742(5) - 1.49(1)/L$ for the
maximum $\partial\mean{|m|}/\partial K$ and $0.8725(3) - 1.50(1)/L$
for the maximum $\partial\log\chi/\partial K$. We conclude that our
rescaled coupling $\kappa$ is the correct scaling, and each of the
effective critical points scale with the same exponent. 

\section{The low and high-temperature regions}
Next we estimate the location of the point where the magnetisation
distribution switches from unimodal to bimodal, denoted
$\kappa^*_c(L)$.  This point will be seen as the effective finite size
separator between the high and low-temperature regions of the model.

\begin{figure}
  \begin{center}
    \includegraphics[width=0.483\textwidth]{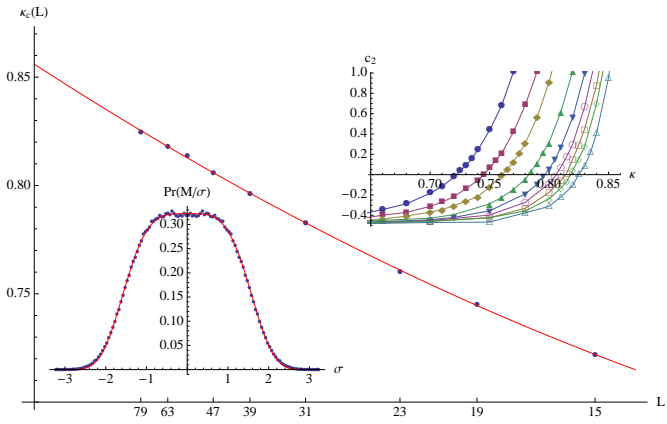}
  \end{center}
  \caption{(Colour on-line) Effective critical temperature
    $\kappa^*_c(L)$ (where $c_2=0$) versus $1/L$ for $L=15$, $19$,
    $23$, $31$, $39$, $47$, $55$, $63$, $79$. The fitted curve is
    $0.856 - 2.52x+7.6x^2$ where $x=1/L$. Lower inset: Probability
    density function $\Pr(M/\sigma)$ (dots) for $L=23$ at
    $\kappa=0.76$ ($K=0.11535165$). Fitted (red) curve is $0.3223
    \exp(-0.002735 x^2 - 0.1164 x^4)$. Upper inset: Coefficient $c_2$
    of the fitted density function $f(x)$ plotted versus $\kappa$ for
    $L=15$, $19$, $23$, $31$, $39$, $47$, $55$, $63$, $79$ (left to
    right).}\label{fig:kcL}
\end{figure}

To do this we note that near this point the standardised distribution
of $M$ (i.e. of $M/\sigma$ where $\sigma^2=\var{M}$) is well-fitted by
the simple formula $f(x)=c_0 \exp(c_2 x^2 + c_4 x^4)$. The lower inset
of Figure~\ref{fig:kcL} shows the distribution of $M/\sigma$ for
$L=23$ at $\kappa=0.76$ ($K=0.11535165$) together with the fitted
formula. The binning of the data was done with Mathematica's built-in
procedures but the results are not very sensitive to this choice. At
this $\kappa$ (i.e. for $L=23$) the distribution is very close to
shifting from unimodal to bimodal, as is indicated by the very small
coefficient $c_2=-0.002735$. The distribution has kurtosis $2.2217$
and when $c_2=0$ the kurtosis becomes $\Gamma(1/4)^4/8\pi^2\approx
2.1884$. In fact, it works equally well to simply measure the kurtosis
and then estimate through interpolation at which point the kurtosis is
$2.1884$.  All this suggests that the distribution's shape is well
captured by $f(x)$. See also \cite{lundow:15a} where we apply this
formula to 5D grids with periodic boundary conditions.

The upper inset of Figure~\ref{fig:kcL} shows the measured
coefficient $c_2$ plotted against $\kappa$ for a range of $L$. The
value of $c_4$ (not shown) at $\kappa^*_c(L)$ is somewhat noisy but it
seems to converge to $-0.114(3)$.  For a cyclic boundary this value is
close to $-0.602$~\cite{pqpaper2}.

Figure~\ref{fig:kcL} shows $\kappa^*_c(L)$ (where $c_2=0$) versus
$1/L$. The fitted 2nd degree polynomial suggests $\kappa^*_c(L) =
0.8561(6) - 2.52(3)/L + 7.6(4)/L^2$, with error bars obtained as
above, so that $\kappa^*_c\approx 0.856$. This point then represents
where the true low-temperature region begins.

The point $\kappa^*_c(L)$ constitutes an effective critical
temperature but its scaling rule appears to depend strongly on the
boundary conditions. Note that for periodic boundary conditions the
corresponding point $K^*_c(L)$ converges to $K_c$ with rate $1/L^3$
\cite{brezin:85}, i.e. faster than the width of the scaling window,
which is $1/L^{5/2}$ in that case. For free boundary conditions we
instead see that $K^*_c(L)\to K_c$ with the {\emph same} rate as the
width of the scaling window, i.e. $1/L^2$.

We note that in \cite{WY2014} an estimate based in the kurtosis was
used, but instead of using the universal kurtosis-value given at the
point where $c_2=0$ they hand picked a value for the kurtosis and then
used the fact that all fixed such values should give the same scaling,
with larger or smaller corrections to the scaling.

\section{Scaling of susceptibility}
\begin{figure}
  \begin{center}
    \includegraphics[width=0.483\textwidth]{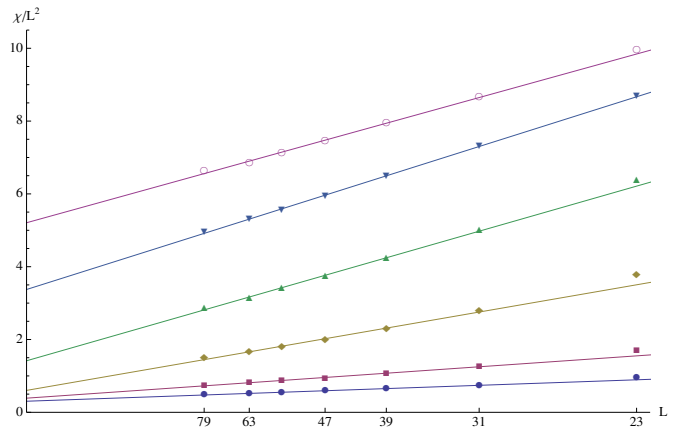}
  \end{center}
  \caption{(Colour on-line) Normalised susceptibility $\chi/L^2$ at
    fixed $\kappa$ versus $1/L$ for $L=23$, $31$, $39$, $47$, $55$,
    $63$, $79$ at $\kappa=0.70$, $0.75$, $0.80$, $0.83$, $0.85$,
    $0.86$ (upwards). The lines are fitted on $L\ge 31$. Error bars
    are smaller than the points.}\label{fig:chiL2a}
\end{figure}

Let us now return to the matter of the finite-size scaling of $\chi$
and $\bar\chi$. With the onset of the low-temperature region at
(asymptotically) $\kappa^*_c=0.856$ we look at the scaling behaviour
of $\chi$ at fixed $\kappa$ below and above this point.  Consider
Figure~\ref{fig:chiL2a} where we show $\chi/L^2$ versus $L$ for six
different fixed $\kappa$-values. Lines are fitted on $L\ge 31$ to
demonstrate the presence of correction terms.  The behaviour is
clearly linear for larger $L$, with the possible exception of
$\kappa=0.86$, i.e. above $\kappa^*_c=0.856$ where we have entered the
low-temperature region and have a bimodal magnetization, but even here
the correction term is still rather weak. However, higher
$\kappa$-values render us stronger corrections.  The normalised
susceptibility was also used in \cite{WY2014}, but there the
estimates, which lead to scaling proportional to $L^{5/2}$ were based
on much smaller lattices $L\leq 36$, where finite size effects are
much stronger.

\begin{figure}
  \begin{center}
    \includegraphics[width=0.483\textwidth]{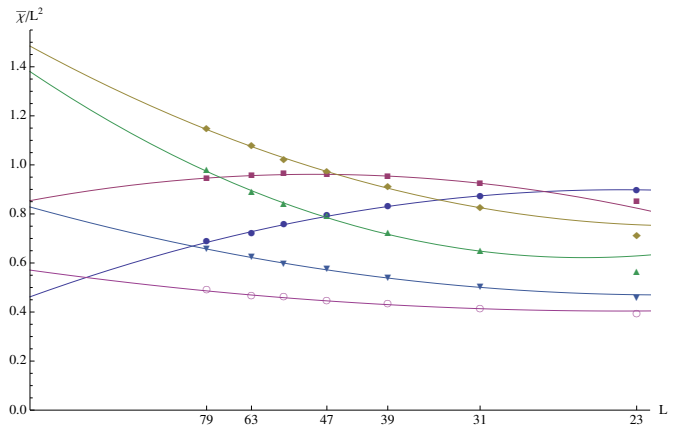}
  \end{center}
  \caption{(Colour on-line) Normalised modulus susceptibility
    $\bar\chi/L^2$ at fixed $\kappa$ versus $1/L$ for $L=23$, $31$,
    $39$, $47$, $55$, $63$, $79$ at $\kappa=0.82$, $0.84$, $0.86$,
    $0.88$, $0.90$ and $0.92$. Downwards at $y$-axis is $\kappa=0.86$,
    $0.88$, $0.84$, $0.90$, $0.92$, $0.82$. The 2nd degree curves are
    fitted on $L\ge 31$. Error bars are smaller than the
    points. }\label{fig:barchiL2a}
\end{figure}

\begin{figure}
  \begin{center}
    \includegraphics[width=0.483\textwidth]{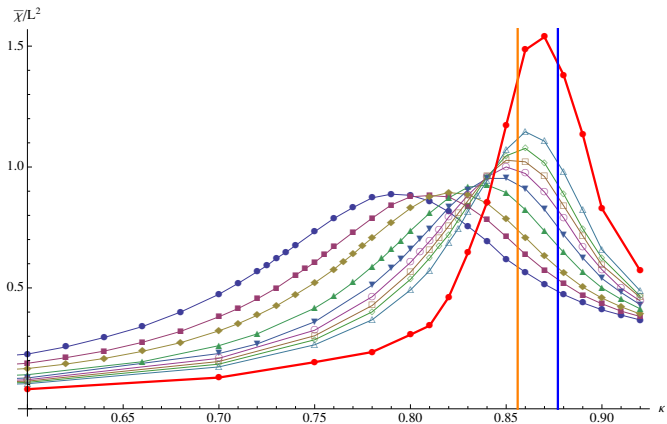}
  \end{center}
  \caption{(Colour on-line) Normalised modulus susceptibility
    $\bar\chi/L^2$ versus $\kappa$ for $L=15$, $19$, $23$, $31$, $39$,
    $47$, $55$, $63$, $79$. The thick red curve is the estimated limit
    found by fitting a 2nd degree polynomial in $1/L$ to the data
    points for $L\ge 31$ at each fixed $\kappa$.  The left vertical
    line at $\kappa=0.856$ (orange) indicates where the magnetisation
    distribution becomes bimodal when $L\to\infty$. The right vertical
    line at $\kappa=0.877$ (blue) is where the maximum $\bar\chi/L^2$
    is located when $L\to\infty$.}\label{fig:barchiL2vk}
\end{figure}

Moving on to the modulus susceptibility $\bar\chi/L^2$ the corrections
become clear and present, even for $\kappa\le 0.856$, but never more
than can be captured by a simple 2nd degree polynomial.
Figure~\ref{fig:barchiL2a} shows this effect. Note also that once we
have gone beyond the maximum at $\kappa=0.876$ the behavior quickly
becomes linear again as is clearly seen at $\kappa=0.92$, which is
where our data end.  Combining all the measured $\bar\chi/L^2$ with
their asymptotic values, based on 2nd degree polynomials fitted on
$L\ge 31$, we now obtain Figure~\ref{fig:barchiL2vk}.

For this range of lattice sizes and at least some values of $\kappa$
one can make a reasonable fit to the data with a $L^{5/2}$ scaling as
well, by including sufficient correction terms. But, as we can see in
Figure~\ref{fig:barchiL2vk}, the $L^2$ scaling gives values that
decrease to their limit value for a large range of $\kappa$, possibly
all $\kappa$ below $\kappa^*_c$, making an even larger power of $L$ seem
unlikely here.

\section{The FK-cluster at the origin}
Our next object of study is the Fortuin-Kasteleyn cluster containing the
origin.  This is the cluster which the Wolff random cluster algorithm
builds when started at the central vertex of our lattice, and is one
of the clusters in the FK-random cluster representation of the Ising
model.  From the FK-model the usual susceptibility can be computed as
the expected cluster size, and with free boundary the expected size of
the origin cluster will be larger than the average cluster size,
thereby giving an upper bound on the susceptibility.  This observation
was used in \cite{lundow:14} to give a susceptibility estimate
which was hoped to be less affected by the effects of the free
boundary, in part as a response to the truncation method of \cite{berche:12}.

\begin{figure}
  \begin{center}
    \includegraphics[width=0.483\textwidth]{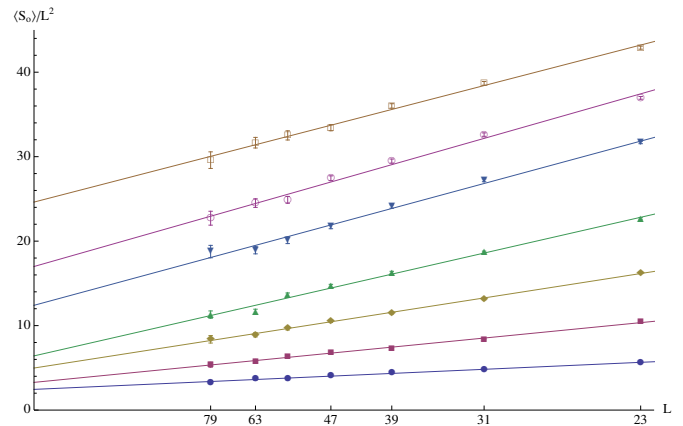}
  \end{center}
  \caption{(Colour on-line) Normalised size of origin cluster $\langle
    S_o\rangle/L^2$ at fixed $\kappa$ versus $1/L$ for $L=23$, $31$,
    $39$, $47$, $55$, $63$, $79$ at $\kappa=0.70$, $0.75$, $0.78$,
    $0.80$, $0.82$, $0.83$ and $0.84$ (upwards). Lines fitted on $L\ge
    23$.  }\label{fig:SoL2a}
\end{figure}

As described we expect $\langle S_o\rangle$ to place an upper bound on
$\chi$, but we do not expect them to be of different scaling
orders. Plotting $\langle S_o\rangle/L^2$ versus $1/L$ demonstrates a
similar behaviour to that of $\chi/L^2$.  In Figure~\ref{fig:SoL2a} we
show the scaling at a number of different $\kappa$-values below
$\kappa^*_c$. In fact, a strong correction enters the picture already
at $\kappa=0.85$ (not shown in the figure). This is to be expected
since above $\kappa^*_c$ there are additional terms from the FK-model
affecting the Ising susceptibility, and those are not included in our
sampled data.

The distribution of the origin cluster size $S_o$ is also an
interesting object.  It is a Pareto-like distribution with density
proportional to $x^{-1-1/\delta}$, or, by definition
\cite{grimmett2004random},
\begin{equation}
  \Pr(S_o\le x) = 1-x^{-1/\delta}
\end{equation}
where $\delta$ is the critical exponent reflecting how an external
field affects the magnetisation at the critical temperature, here
having the mean field value $\delta=3$.

\begin{figure}
  \begin{center}
    \includegraphics[width=0.483\textwidth]{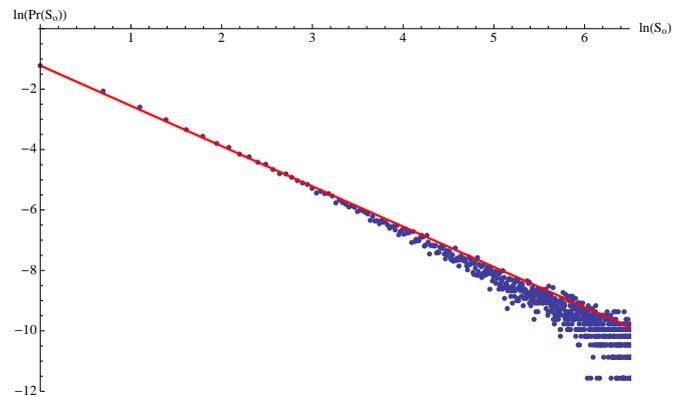}
  \end{center}
  \caption{(Colour on-line) Log-log plot of the size distribution of
    the origin cluster for $L=63$ and $\kappa=0.85$.  The red line has
    slope $-4/3$. }\label{fig:Sodist1}
\end{figure}

In Figure~\ref{fig:Sodist1} we show a log-log plot of the distribution
density (for small $x=S_o$) together with a line having slope $-4/3$,
consistent with $\delta=3$. The line is clearly an excellent fit.  The
plot shows the distribution for $L=63$ at $\kappa=0.85$ but for this
range of cluster sizes the plot is indistinguishable from that of
other $\kappa$-values, and indeed of other $L$ (except the smallest
$L$).

\section{Correlation length}
The origin cluster can also be used to estimate the correlation length
of the model.  In the high temperature region the correlation length
$\xi$ is given, see \cite{grimmett2004random}, by the limit
\begin{equation}\label{corr}
  \frac{1}{\xi} = \lim_{L\to\infty} \frac{-\log\Pr(o\leftrightarrow\partial G)}{L}
\end{equation}
For a fixed $\kappa$ we expect not just $\xi\to\infty$, but rather
that the correlation length is comparable to $L$, i.e. $\xi/L\to
a(\kappa)$ for some function $a$.  Thus we propose to estimate
$a(\kappa)$ by first defining $\xi/L=-1/\log\Pr(o\leftrightarrow
\partial G)$ and then use a simple projection rule to estimate the
limit function $a(\kappa)$.

\begin{figure}
  \begin{center}
    \includegraphics[width=0.483\textwidth]{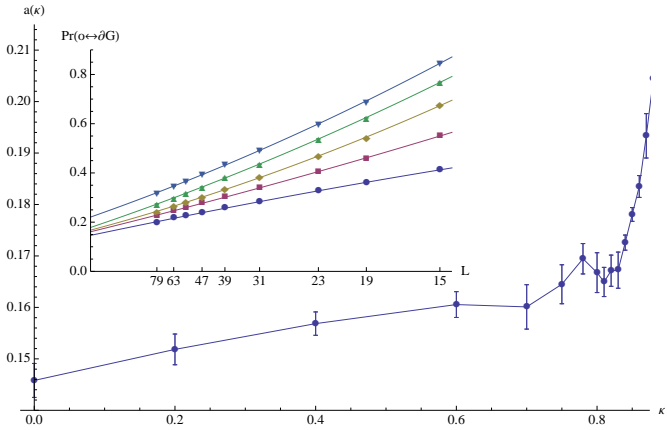}
  \end{center}
  \caption{(Colour on-line) Correlation coefficient $a(\kappa)$ versus
    $\kappa$ based on fitted 2nd degree polynomials for $L\ge 15$ (see
    text). Note the sudden growth that sets in near $\kappa=0.85$.
    Inset: $\xi/L$ plotted against $1/L$ for $L=23$, $31$, $39$, $47$,
    $55$, $63$ and $79$ at at $\kappa=0$, $0.40$, $0.70$, $0.80$,
    $0.84$, $0.88$ and $0.92$ (upwards). Curves are 2nd degree
    polynomials fitted to $15\le L\le 79$.  }\label{fig:xiLavk}
\end{figure}

The inset of Figure~\ref{fig:xiLavk} shows $\xi/L$ versus $1/L$ for a
range of $L$ at different $\kappa$-values. Figure~\ref{fig:xiLavk}
also shows a rough estimate of $a(\kappa)$ based upon fitted 2nd
degree polynomials (on $L\ge 15$) for each $\kappa$. We expect the
errors in the plot to be significant although the general behaviour as
such is correct. In the low-temperature region the identity in
Equation \ref{corr} is no longer valid, and the probability that the
origin cluster reaches the boundary tends to 1, and this agrees well
with the steep growth beginning at $\kappa\approx 0.85$, the start of
the effective finite size low-temperature region.

\section{Conclusions}
As we have seen here our data fits excellently with the conventional
scaling picture for the free boundary case.  Now, apart from this fact
we believe that there are good reasons for expecting distinct
behaviour from the free and cyclic boundary cases.  As is well known
the Ising model is equivalent to the Fortuin-Kasteleyn random cluster
model for $q=2$. The case $q=1$ corresponds to ordinary percolation
and $q\rightarrow 0$ gives the random spanning tree for the
lattice. 

For the random spanning tree Pemantle \cite{MR1127715} proved
that for large enough $d$ the distance between two points in the tree
scales as $L^2$ with free boundary and as $L^{d/2}$ with cyclic
boundary.  For percolation, Aizenmann \cite{aizenman:97} conjectured
that for large enough $d$ the size of the largest cluster should scale
as $L^4$ with free boundary and $L^{2d/3}$ with cyclic boundary. This
conjecture was later proven in \cite{HH:07,HH:11}. So, for both $q=0$
and $q=1$ we have rigorous results showing that free and cyclic
boundary conditions lead to distinct scalings. 

Aizenman's conjectured scaling $L^{2d/3}$ came from the fact that the
cyclic boundary case was expected to behave like the critical Erd\H
os-Renyi random graph, where the maximum cluster has size $N^{2/3}$,
where $N$ is the number of vertices, which would be $L^d$ for the
lattice.  Again, there are detailed rigorous results \cite{luczak:06}
concerning the random cluster model on the complete graph and in
\cite{lundow:15b} it was demonstrated that Monte Carlo data for the
FK-model with $q=2$ and $d=5$ with cyclic boundary gives a detailed
agreement with the scaling for the complete graph.  In particular, the
size of the largest cluster has the same scaling for both graphs for several
different rescaled coupling ranges.

Now, the $L^2$ scaling observed in this paper for the origin cluster
can be seen as an indicator of the correct scaling of the largest
cluster as well. This would give a scaling which is different for the
free and cyclic boundary cases, just as one would expect from the
known results for lower $q$.  In \cite{boundarypaper} the authors
conjectured that for every $q$ there is a $d(q)$ such that above this
dimension the FK-model with cyclic boundary behaves like the model on
a complete graph, giving a partial generalisation of Aizenmann's
conjecture for all $q$.  In fact, we also expect the model with free
boundary to follow a simpler mean-field scaling, similar to that seen
on an infinite tree, or Bethe lattice.  The fact that there are
several distinct basic model systems, e.g. the complete graphs and the
Bethe lattices, which on one hand have mean field critical exponents
but on the other hand differ on more detailed properties is probably
under-appreciated in the older literature.

\begin{acknowledgments}
  The simulations were performed on resources provided by the Swedish
  National Infrastructure for Computing (SNIC) at High Performance
  Computing Center North (HPC2N) and at Chalmers Centre for
  Computational Science and Engineering (C3SE).
\end{acknowledgments}


\end{document}